\documentclass[twocolumn,preprintnumbers,nopacs,amsmath,amssymb]{revtex4}
\usepackage{graphicx}
\usepackage{dcolumn}
\usepackage{bm}

\begin{document}
\title{The theory of quantum levitators}
\author{Fran\c{c}ois Impens$^1$, Franck Pereira Dos Santos$^2$,
Christian J. Bord\'e$^2$}
\address{$^1$ Instituto de F\'{\i}sica, Universidade Federal do Rio de Janeiro. Caixa Postal 68528, 21941-972 Rio de Janeiro, RJ, Brazil.}
\address{$^2$ LNE-SYRTE, Observatoire de Paris, LNE, CNRS, UPMC; 61 avenue de l'Observatoire, 75014 Paris, France.}
\begin{abstract}
 We develop a unified theory for clocks and gravimeters using the interferences of multiple atomic waves put in levitation by traveling light pulses. Inspired by optical methods, we exhibit a propagation invariant, which enables to derive analytically the wave function of the sample scattering on the light pulse sequence. A complete characterization of the device sensitivity with respect to frequency or to acceleration measurements is obtained. These results agree with previous numerical simulations and confirm the conjecture of sensitivity improvement through multiple atomic wave interferences. A realistic experimental implementation for such clock architecture is discussed.
\end{abstract}
\maketitle

 Atom interferometers~\cite{Berman} have been used successfully to build accurate optical clocks~\cite{Wilpers07} and gravimeters~\cite{Chu99,Malossi10}. However, in the presence of a gravity field, these systems faced critical limitations caused by the free fall of the atoms used in the interferometric measurement. Among these, we mention the finite time of atomic interrogation, and the curvature shifts resulting from the transverse field exploration by the traveling atoms~\cite{Trebst01}.

Recently, we have exposed a strategy to overcome these limits, which relies on the atomic levitation thanks to a series of vertical traveling light pulses~\cite{Impens06,Impens09b}. A unique laser field performs simultaneously the atomic confinement and the atomic interrogation, which are usually operated by  distinct laser fields in typical setups~\cite{BestOpticalClocks}. In our initial proposal~\cite{Impens06}, the setup juggles with a single atomic cloud thanks to a series of synchronized velocity-sensitive Raman $\pi$-pulses. The feasibility of this strategy was demonstrated by Hughes et al.~\cite{Hughes09} in a similar experiment achieving over a hundred bounces. In a second proposal~\cite{Impens09b}, we considered a different atomic sustentation process, which exploits instead the quantum interferences of multiple flying wave-packets. These packets bounce on a series of short $\pi/2$-pulses, which give rise to a myriad of trajectories and yield a multiple-wave atom interferometer~\cite{MultipleWaveAtomInterferometry} in levitation. The phases of the light pulses are finely tuned in order to induce constructive interferences in a network of preferred levitating paths. An analog of such quantum trampoline has been realized recently by Robert-de-Saint-Vincent et al.~\cite{SaintVincent10}.\\

A moving two-level atom can be considered as a clock, and significant advances can be achieved by adopting a unified point of view for clock and inertial sensors~\cite{BordeABCD}. This is exactly the philosophy of our approach, which gives a complete quantum description of the levitating atoms. Preliminary steps towards the construction of a levitating atomic gravimeter have been taken~\cite{Hughes09,SaintVincent10}. In contrast, levitating atomic clocks still await their experimental realization. The purpose of this article is to explore in greater detail the behavior and the sensitivity of such suspended atomic systems. We also bring out suitable atomic species for these setups, preparing the way towards the implementation of such clock architectures.\\

We proceed as follows. Sec.\ref{sec:twopulse} presents the theory of the atomic phase-matching with two separated $\pi/2$ light pulses. We expose in greater detail the argument of Ref.\cite{Impens09b} and derive the propagation invariant used in the subsequent developments. Sec.\ref{sec:levitation} recalls the principle of the atomic sensor using multiple-wave atomic levitation. In Sec.\ref{sec:short resonant light pulses}, we derive the levitating wave-function when the system operates at resonance. In Sec.\ref{sec:nonresonant short pulse regime}, we establish the levitating wave-function when the system operates out of resonance, and obtain the sensor sensitivity as a clock or as a gravimeter in the short-pulse regime. We obtain analytical expressions for the fundamental sensitivity limit of such systems, in remarkable agreement with previous numerical simulations presented in~\cite{Impens06,Impens09b}. In Sec.\ref{sec:clock systems}, we exhibit atomic systems suitable for this setup with realistic experimental parameters.

\section{Atomic phase-matching with a sequence of $\pi/2$-pulses.}
\label{sec:twopulse}

\begin{figure}[htbp]
\begin{center}
\includegraphics[width=8cm]{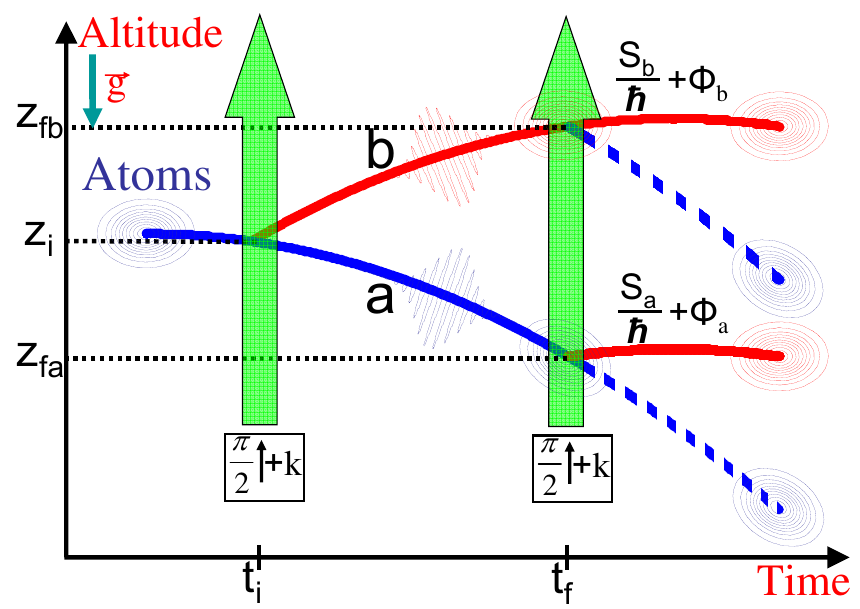}
\end{center}
\caption{Action of a pair of copropagating $\pi/2$-pulses on a
free-falling atomic wave-packet. The phase difference between the
two outgoing wave-packets comes from the classical action and
laser phase acquired on each path, and from the distance of their
centers.} \label{fig:sequence deux pisurdeux}
\end{figure}

We consider a dilute two-level atomic sample well-described by a two-component wave-function, noted $(\psi_b(z,t),\psi_a(z,t))$, and initially entirely in one of these two energy levels. This sample experiences a free fall in the uniform gravity field, thus following the Hamiltonian
\begin{equation}
\hat{H}=  \frac {\hat{p}^2} {2 m} + m g \hat{z}+ E_b | b \rangle \langle b |+ E_a | a \rangle \langle a | \, .
\end{equation}
A key idea of the levitating multiple-wave interferometer is the fine-tuning of atomic phases through a sequence of $\pi/2$-pulse pairs. In order to expose this concept, it is useful to give a short reminder on the propagation of ideal atomic waves in time-dependent quadratic potentials with the $ABCD$ method~\cite{BordeABCD}. Since the external potential is one-dimensional, we restrict without loss of generality the discussion to 1D wave-packets, taking
an initial sample wave-function $\psi_l(z,t_0)= wp_l(z,z_{c0},p_{c0},w_0)$ of the form
\begin{equation}
\label{eq:twopulse initial wave packet 1D Gaussian position}
wp_l(z,z_{c0},p_{c0},w_0)  =  \frac {1} {\sqrt{\pi \: w_0}} e^{- (z-z_{c0})^2 /2 w_0^{2}  + i  p_{c0} (z-z_{c0}) / \hbar}
 \end{equation}
Using the $ABCD$ theorem for atomic waves~\cite{BordeABCD}, one can express
this wave-packet at a later instant $t>t_0$ as
\begin{equation}
\label{eq:twopulse wave packet 1D Gaussian position}
\psi_l(z,t)  = wp_l(z,z_{c}(t),p_{c}(t),w(t)) e^{i S_l(z_{c0},p_{c0},t-t_0)/\hbar}
 \end{equation}
The central position $z_c(t)$ and central momentum $p_c(t)$ of the wave-packet follow the classical equations of motion of a point-like particle evolving in the considered Hamiltonian, situated at the position $z_{c0}$ with the momentum $p_{c0}$ at the instant $t=t_0$. The term $S_l(z_{c0},p_{c0},t-t_0)$ is the action of the particle in the internal state $l$, integrated along the classical trajectory
\begin{equation}
S_l(z_{c0},p_{c0},t-t_0)= \int_{t_0}^{t} dt' \left(  \frac {p^2_c(t')} {2 m} - E_{l}- m g z_c(t') \right)
\end{equation}
The width $w$ is simply given by
\begin{equation}
w(t)=\sqrt{w_{0}^2+ \frac {\hbar^2} {m^2 w_{0}^2} (t-t_0)^2   }
\end{equation}
It is useful to express the $ABCD$ theorem in the momentum picture. The Fourier transform of the wave-packet of Eq.~\eqref{eq:twopulse initial wave packet 1D Gaussian position} gives readily
 \begin{equation}
\label{eq:twopulse paquet initial impulsion} \psi_l(p,t_0)=  \sqrt{\frac {1} {\pi \: w_{p 0}}}  e^{-
 \frac {(p-p_{c 0})^2 } {2 w_{p 0}^2} } e^{-  i z_{c 0} p /\hbar}
\end{equation}
Performing a Fourier transform on the packet of  Eq.\eqref{eq:twopulse wave packet 1D Gaussian position}, one immediately obtains the momentum wave-packet at a later instant
\begin{equation}
\label{eq:twopulse paquet final impulsion} \psi_l(p,t)=  \sqrt{\frac {1} {\pi \: w_p}}  e^{-
 \frac {(p-p_{c})^2} {2 w_p^2} } e^{i S_l(z_{c0},p_{c0},t-t_0)/\hbar} e^{-  i  z_{c} p / \hbar}
\end{equation}
We have introduced the initial and final width in momentum  $w_{p 0}= \hbar/ w_0$ and $w_{p}= \hbar/ w$. The effect of a short $\pi/2$-pulse of wave vector $\mathbf{k}= \: \epsilon \:  k \: \mathbf{z}$
\footnotemark[1] \footnotetext[1]{We adopt the convention that $k>0$ is the absolute value of the wave-vector associated with all the considered pulses, and we use $\epsilon=1$ and $\epsilon=-1$ for an upward and a downward-traveling pulse respectively - the axis $O_z$ is oriented upwards -. } on an atomic wave-packet is efficiently described by a $2 \times 2$ Rabi matrix evaluated at the central instant $t_c$ of the pulse
\begin{equation}
\label{eq:twopulse matrice Rabi pisur2}
\frac {1} {\sqrt{2}} \left(\begin{array} {cc} 1  & i  \: e^{-i (\epsilon k  \hat{z}-\omega t)}\\
i \: e^{i ( \epsilon k  \hat{z}-\omega t)} & 1 \\
\end{array} \right) \,.
\end{equation}
In this expression $\hat{z}$ is a position operator. The non-diagonal matrix elements are thus quantum operators involving the laser phase $\phi(\hat{z},t)= \epsilon k  \hat{z}-\omega t$, with $\epsilon=1$ and $\epsilon=-1$ for an upward and downward-traveling pulse respectively. Their action onto an atomic wave-packet reflects the acquisition of a laser phase $\phi(z_c,t_c)$ and of a momentum quantum through the atomic recoil
\begin{eqnarray}
e^{i ( \epsilon k  \hat{z}-\omega t_c)} [wp_l(z,z_{c},p_{c},w)] &  = &  e^{i ( \epsilon k  z_c-\omega t_c)} \\
& \times & wp_l(z,z_{c},p_{c}+\epsilon  \hbar k,w) \,. \nonumber
\end{eqnarray}
The proposed experiment rests on the ability to perform a controlled momentum transfer from the light field to the atoms thanks to a sequence of $\pi/2$-pulses. In this view, it is important to understand how such transfer can be done with two $\pi/2$-pulses separated by a time interval. There is an obvious similarity with the Ramsey interrogation, but the novelty of our approach is that it takes into account the quantized atomic motion in the gravitational field. Fig.\ref{fig:sequence deux pisurdeux} illustrates the effect of this pulse sequence on an atomic sample.
 We consider two short upward $\pi/2$-pulses performed respectively at the times $t_i=0$ and $t_f=T$ on the sample and with the phase expressed with the central time as reference:
  \begin{equation}
  \phi_1(z,t)= k z - \omega_1 (t-T/2), \quad \phi_2(z,t)= k z - \omega_2 (t-T/2)+\phi_c \, . \label{eq:laser pulse phases}
   \end{equation}
   Note that the evaluation of the phase at the central time is an exact result, which follows from the $ttt$ theorem~\cite{AntoineTTT}, and not a mere approximation. The gravitational acceleration yields different resonance conditions for the first and the second light pulse, to which the frequencies $\omega_1$ and $\omega_2$ are adjusted respectively. The pulses split the atomic cloud into four wave-packets, following two intermediate paths labeled with the internal atomic states $a,b$. The upper-state and lower-state wave-functions  receive a contribution from each path, corresponding to wave-packets of different central altitude but of common central momentum. In order to describe the quantum interferences between the atomic wave-packets after this sequence, it is thus easier to work in the momentum picture, for which the packets of a given energy share the same center. We consider an initial lower-state wave-packet
   \begin{equation}
\label{eq:twopulse initial double wave vector}
\psi(z,t_0)=  \sqrt{\frac {w_0} {\pi \: \hbar}} \left(\begin{array} {c}
  0 \\
 e^{-
 \frac {(p-p_{c 0})^2 } {2 w_p^2} } e^{-  i z_{c 0} (p-p_{c0}) /\hbar} \\
\end{array}
\right)
\end{equation}
of a similar form as Eq.\eqref{eq:twopulse paquet initial impulsion} up to a different constant phase. This choice will turn out convenient to compute the levitating wave-function. Using the Rabi matrix of Eq.\eqref{eq:twopulse matrice Rabi pisur2}, and the $ABCD$ law in momentum picture given by the Eqs.(\ref{eq:twopulse paquet initial impulsion},\ref{eq:twopulse paquet final impulsion}), one readily obtains the form of the wave-function after two $\pi/2$-pulses
\begin{equation}
\psi(z,T)= \frac {1} {2} \left(\begin{array} {c}
 e^{- \frac {(p-p_{c b})^2}
{2 w_p^2} } \sum_{l=a,b} e^{- \frac i \hbar z_{f l} (p-p_{f b})+ i \Phi_{l,b}}  \\
 e^{- \frac {(p-p_{c a})^2} {
2 w_p^2}} \sum_{l=a,b} e^{- \frac i \hbar z_{f l} (p-p_{f a})+ i \Phi_{l,a}} \\
\end{array}
\right)
\end{equation}
We have noted $z_{f a}$, $p_{f a}=p_0- m g T$ and $z_{f b}$,$p_{f b}=p_0+\hbar k- m g T$) the respective central position and momentum of the lower-state and upper-state atomic packet immediately before the second light pulse~[See Fig.\ref{fig:sequence deux pisurdeux}]. In the equation above, the phase $\Phi_{e_1, e_2}$ is associated with the wave-packet that was in the internal state $e_1$ after the first light pulse and in the internal state $e_2$ after the second pulse. We note $\phi_{e_1, e_2}$ the contribution to this phase arising from the atom-light interaction. The expression of the phases $\Phi_{e_1, e_2}$ follows
from the $ABCD$ theorem in momentum picture and from the Rabi matrix
\begin{equation}
\label{eq:twopulse phases}
\Phi_{e_1,e_2}  =  (p_{0} z_{0}-p_{f e_2} z_{f e_1})/\hbar +S_{e_1}/\hbar+\phi_{e_1, e_2} \,.
\end{equation}
We have used the short-hand notations $S_a=S_a(z_0,p_0,T)$ and $S_b=S_b(z_0,p_0+\hbar k,T)$. The phase $\phi_{e_1, e_2}$ read $\phi_{a,a}=0$, $\phi_{a,b}=\phi_2(z_{f a},T)$, $\phi_{b,a}=\phi_1(z_0,0)-\phi_2(z_{f b},T)$ and $\phi_{b,b}=\phi_1(z_0,0)$. In order to obtain a full momentum transfer for the sample, one aims at producing constructive interferences in the upper atomic state and destructive interferences in the lower atomic state, i.e. the outgoing wave-function $\psi(p,T)$ should be of the form
\begin{equation}
\left(\begin{array} {c}
 i e^{- \frac {(p-p_{f b})^2} {2 w_p^2}+i \Phi_{f b}} \left( e^{- \frac i \hbar z_{f a} (p-p_1)}+ e^{- \frac i \hbar z_{f b} (p-p_1)} \right)  \\
 e^{- \frac {(p-p_{f a})^2} {2 w_p^2}+i \Phi_{f a}}  \left( e^{- \frac i \hbar z_{f a} (p-p_0)}- e^{- \frac i \hbar z_{f b} (p-p_0)} \right) \\
\end{array}
\right) \label{eq:twopulse final momentum wavepacket}
\end{equation}
This is possible only if the phases~\eqref{eq:twopulse phases} satisfy the relations
\begin{equation}
\Phi_{a,b}=\Phi_{b,b}, \quad  \Phi_{b,a}=\Phi_{a,a}+\pi
\end{equation}
 In order to preserve the norm of the double component wave-function $\psi$ during the unitary evolution, these relations are not independent: it is for instance sufficient to ensure constructive interferences in the upper state in order to obtain destructive interferences in the lower state. We thus choose without loss of generality to satisfy the condition $\Phi_{a,b}=\Phi_{b,b}$.
 Considering the phase reference used for the pulses~[see Eq.\eqref{eq:laser pulse phases}], one sees that the previous relation is satisfied if and only $\phi_c=0$ and if the pulse frequencies are adjusted to the values
\begin{equation}
\label{eq:twopulse resonance frequencies} \omega_{1,2}^0 \: = \: \frac {1}
{\hbar} \: \left( E_b+ \frac {(p_{1,2}+ \hbar
k)^2} {2 m} - E_a - \frac {p_{1,2}^2} {2 m} \right)
\end{equation}
with $p_1=p_0$ and $p_2=p_0- m g T$. These equations state simply the energy conservation for each pulse, taking into account both the change of internal level and the atomic recoil. One should note that the phase matching given by the relations~\eqref{eq:twopulse resonance frequencies} is not sufficient to guarantee efficient constructive interferences in the excited state. One must also require a sufficient overlap of the interfering wave-packets, namely $w_p (z_{f b}-z_{f a}) \ll  1$, or equivalently
\begin{equation}
\label{eq:twopulse coherence condition}
\Delta \omega_{\mbox{Doppler}} T \ll 1 
\end{equation}
with $\Delta \omega_{\mbox{Doppler}}=k w_p/m$ the Doppler frequency spread experienced by the atomic sample because of its dispersion in momentum. We note that this condition can also be expressed in terms of Ramsey interrogation as $\Delta \omega_{\mbox{Doppler}} \ll  \Delta \omega_{\mbox{Ramsey}} / (2 \pi)$
since
$\Delta \omega_{\mbox{Ramsey}}= 2 \pi /T$ is the frequency resolution expected from the Ramsey interrogation with the considered sequence of $\pi/2$-pulses. If this condition is not met, the momentum spread reduces the constructive interferences through a partial overlap of the atomic wave-packets. Alternatively, this condition can be stated as $w \gg v_r T$: the vertical coherence length of the atomic wave-packets must be much greater than the separation of their respective centers in order to yield constructive interferences. This is thus a perfectly intuitive result.

We establish now a key property: the phase $\Phi_{a,b}=\Phi_{b,b}$ picked up by the wave-packet ending in the upper state is not only path-independent but also independent from the initial position $z_0$, providing us with an invariant~\footnotemark[1] \footnotetext[1]{We correct here a typal in~\cite{Impens09b}, which explains the difference between Eq.\eqref{eq:twopulse invariant I} and the invariant considered in~\cite{Impens09b}.}
\begin{equation}
\label{eq:twopulse invariant I}
I= p_i z_i - p_f z_f + S + \hbar \phi \,.
\end{equation}
 To show this, we compute the quantity $I$ on each path,
 \begin{eqnarray}
 \label{eq:twopulse Ia}
 I_a(z_i,p_i,T) & = & p_i z_i-p_f (z_i+p_i T/m-gT^2/2)  \\
 & + &S_a(z_i,p_i,T)+\phi_2(z_i+p_iT/m-gT^2,T)  \nonumber  \\
   I_b(z_i,p_i,T) & = & p_i z_i-p_f (z_i+(p_i+\hbar k) T/m-gT^2/2)
   \nonumber \\
   &+& S_b(z_i,p_i+\hbar k,T)+\phi_1(z_i,0) \label{eq:twopulse Ib} \,,
 \end{eqnarray}
when the resonance conditions of Eq.\eqref{eq:twopulse resonance frequencies} and the condition $\phi_c=0$ are satisfied. A straightforward computation gives $I_a(z_i,p_i,T)=I_b(z_i,p_i,T)=I(p_i,k,T)$ with
 \begin{eqnarray}
 \label{eq:twopulse expression invariant I}
 I(k,p_i,T) & = & -\left[ \frac {1} {2}(E_b+E_a) +\frac {p_i^2} {2 m}\right] T \nonumber \\
 & \: & +  \frac 1 2 (p_i+\hbar k) g T^2 - \frac 1 6 m g^2 T^3
 \end{eqnarray}
 At a deeper level, the path-invariance of the quantity $I$ is a consequence of an important feature of atomic-wave propagation, which becomes obvious in a 5D-formalism~\cite{ChristianEPJD08}: the center of the wave-packet follows an equiphase trajectory with a generalized phase including an extra-degree of freedom, namely the proper time conjugated to the atomic mass. As announced, the invariant $I$ depends on the initial momentum but not on the initial position. This property allows one to address simultaneously the levitating atomic packets situated at different altitudes with a single laser frequency. Besides, this fact turns the momentum picture into a very convenient framework for the description of the levitating interferometer.

\section{Principle of the multiple-wave atomic levitation.}
\label{sec:levitation}

\begin{figure}[htbp]
\begin{center}
\includegraphics[width=7cm]{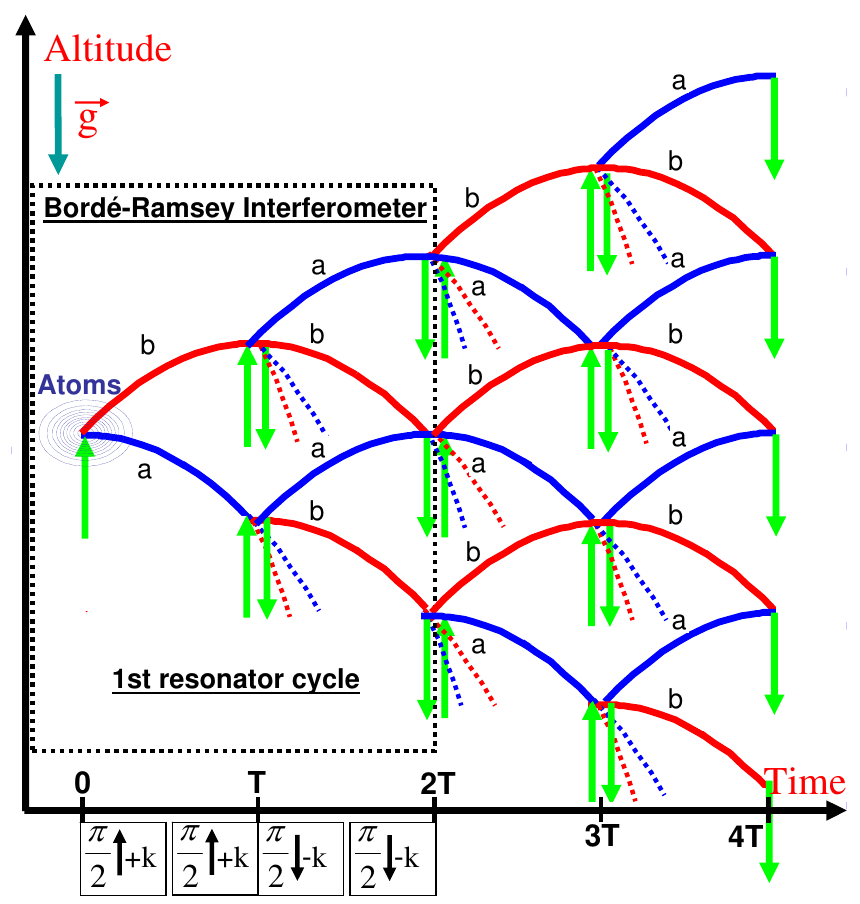}
\end{center}
\caption{Levitating atomic trajectories in the
sequence of pulses. The first four pulses generate a vertical
Bord\'e-Ramsey interferometer. The central positions of the
wave-packets explore a network of paths whose number doubles at each
laser pulse.} \label{fig:levitation arches}
\end{figure}

For convenience, we expose in this Section the principle of the sensor using multiple-wave levitation, following the lines of Ref.\cite{Impens09b}. In order to obtain the desired suspension of the atomic sample, one extends the previous pulse sequence as follows. Immediately after the pair of upward-travelling $\pi/2$-pulses, one performs a pair of downward-travelling $\pi/2$-pulses with an identical interpulse duration. The four-pulse sequence obtained constitutes a Bord\'e-Ramsey interferometer, bent here by the gravitational field. This interferometer corresponds to the trajectories in the dashed box of Fig.~\ref{fig:levitation arches}.
The previous discussion shows that this second sequence performs a nearly full population transfer if the energy conservation is verified for each pulse - using again the mid-time of the pulse pair as phase reference - and if the momentum dispersion is small enough as to fulfill Eq.\eqref{eq:twopulse coherence condition}. One then achieves a net momentum transfer of $\Delta p = 2 \hbar k$ per atom. Besides, for the special interferometer duration
\begin{equation}
\label{eq:levitation resonance T}
T^{0}= \frac {\hbar k} {mg},
\end{equation}
the atoms end up with their initial momentum value. For this special duration, by repeating the four-pulse sequence, one obtains a periodic atomic motion in momentum, thereby enabling levitation if the sample is initially at rest. Thanks to this periodicity, the energy conservation relations~\eqref{eq:twopulse resonance frequencies} are identical for each four-pulse sequence. Indeed, these conditions yield only two different resonant frequencies splitted from the transition frequency $\omega_{ab}=(E_b-E_a)/\hbar $ by the recoil frequency
\begin{equation}
\label{eq:levitation resonance frequencies}
\omega^0_{1,4}= \omega_{ba} + \frac {\hbar k^2} {2 m}, \quad \omega^0_{2,3}= \omega_{ba} - \frac {\hbar k^2} {2 m}
\end{equation}
The first and the fourth pulse, as well as the second and the third ones, have identical resonant frequencies.

When the previous resonance conditions ($T:=T_0,\omega_{1,2,3,4}:=\omega^0_{1,2,3,4}$) are fulfilled, and for a sufficiently narrow atomic momentum
distribution~\eqref{eq:twopulse coherence condition}, the atomic sample spreads through the network of suspended trajectories sketched on Fig.~\ref{fig:levitation arches}. As explained in Ref.\cite{Impens09b}, when the parameters $(T,\omega_{1,2,3,4})$ differ from these resonant values, the phase-matching condition is violated either because of the velocity drift if $T \neq T_0$, either because of the mismatch in the pulse frequencies. This opens quantum channels out of the levitating paths and induce losses in the suspended sample. By measuring the sample population after a given number of cycles with resonant frequencies $\omega_{1,2,3,4}:= \omega^0_{1,2,3,4}$  and for different values of the period $T$, one can thus determine $T_0$ and measure the local gravitational acceleration through Eq.\eqref{eq:levitation resonance T}. This measurement indeed also rests on the
determination of the recoil velocity $v_r= \hbar k/ m$, but the latter quantity is usually known with an excellent accuracy. It is worth noticing that the recoil velocity may also be determined independently by the present experiment. When the period $T$ is fixed to $T:=T_0$, the suspended cloud looses atoms if the pulse frequencies $\omega_{1,2,3,4}$ are shifted from their resonant values. These mechanism can be exploited to lock the lasers performing the suspension on these resonant frequencies.

\begin{figure}[htbp]
\begin{center}
\includegraphics[width=8cm]{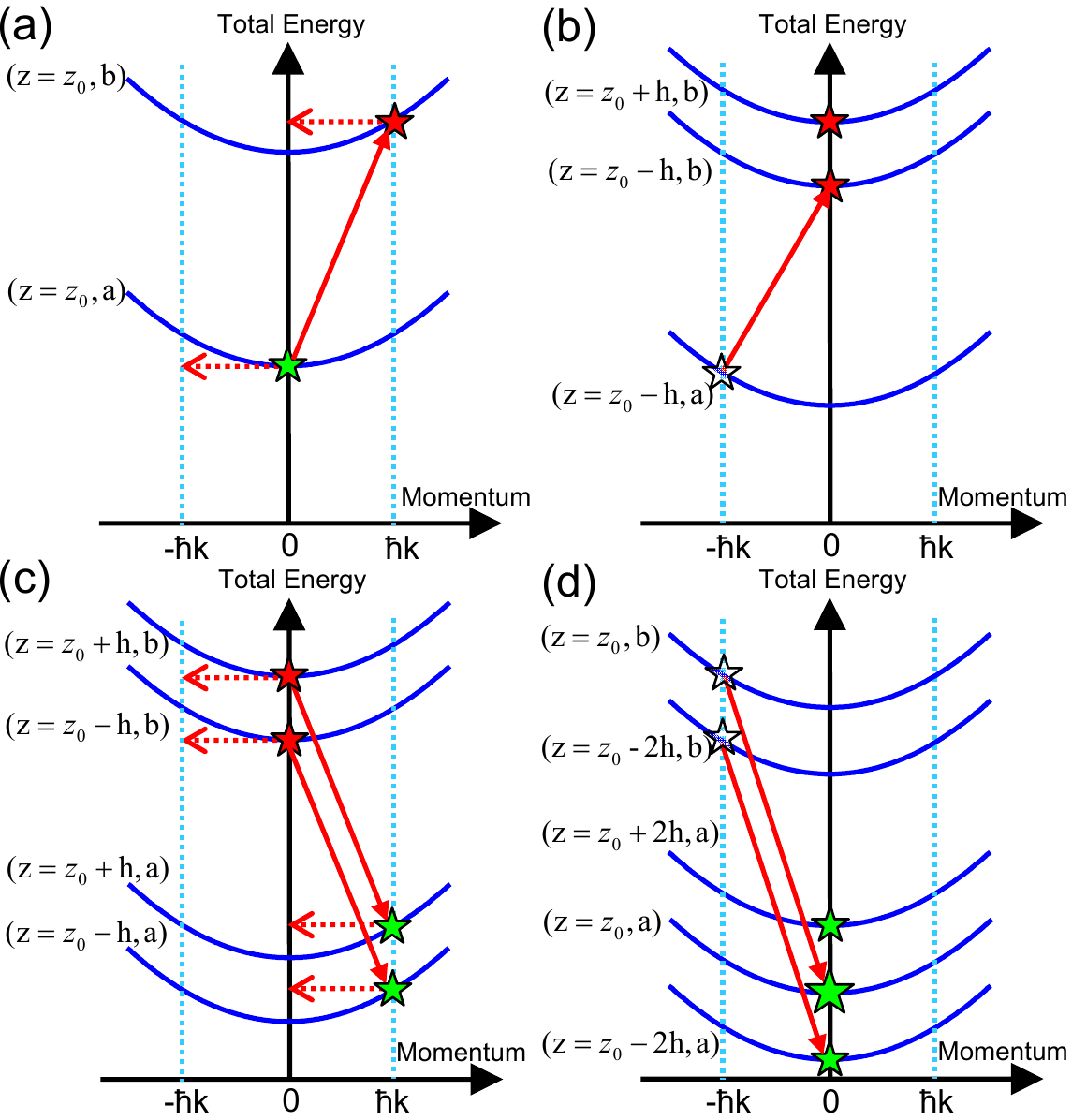}
\end{center}
\caption{Motion of the atomic wave-packets in the
energy-momentum picture for the interferometer duration $2 T_0$.
Figs.~(a,b,c,d), associated with the 1st,2nd,3rd and 4th light
pulses respectively, show the packets present in coherent
superposition (full stars) immediately after - or transferred
(transparent stars) during - the considered pulse, whose effect is
represented by a full red arrow.} \label{fig:energy momentum}
\end{figure}

We recall here the discussion of Ref.\cite{Impens09b} showing that the proposed setup is analog to an atomic resonator in the momentum picture. It is insightful to consider, as plotted on Fig.\ref{fig:energy momentum}, the motion of the atomic wave-packets in the energy-momentum picture when both resonance conditions are satisfied. The energy includes contributions from the internal level energy, from the kinetic energy and from the gravitational potential energy. For the non-relativistic motion considered here, the energy of an atomic packet is a quadratic function of its average momentum. The energy-momentum curves are thus a family of parabolas, labeled with the internal level and with the average altitude of the atomic packet. Each star on the Fig.\ref{fig:energy momentum} stands for an atomic wave-packet. Since the atomic motion is conservative and accelerated downwards, these stars follow an horizontal leftward trajectory. Fig.\ref{fig:energy momentum} reveals that, in the momentum picture, the atomic motion is periodic and occurs in a region bounded by two well-defined values $\pm \hbar k$ corresponding to the atomic recoil. This property strongly suggests an analogy with a momentum resonator. Momentum confinement is provided here by destructive interferences which shut off the quantum channels going out of the considered region.

\section{Scattering on a sequence of short resonant light pulses.}
\label{sec:short resonant light pulses}
As a preliminary step toward the prediction of the interferometer fringe pattern, we compute the macroscopic wave-function associated with the levitating atoms in a sequence of resonant $\pi/2$ pulses. The transverse degrees of freedom play a trivial role in the considered setup, so that one can follow without loss of generality a one-dimensional approach. In the following, the time-dependence of the momentum width is given by $w_p(t)=\hbar/\sqrt{w_{0}^2+ \hbar^2 (t-t_0)^2 / w_{0}^2  }$. It will be simply noted $w_p$.

We shall use the following property - proven in Appendix -. After $N$ pairs of $\pi/2$-pulses, i.e. at the time $t_{N}=N T$, the wave-function associated with the levitating atoms is of the form
\begin{eqnarray}
\label{eq:sommeresonante1}
\psi_{\mbox{lev}}(p,t_{N}) & = & \frac {1} {2^{N} \pi^{1/4} \sqrt {w_p}} e^{i \Phi_N} e^{-
\frac {(p-p_N)^2} {2 w_p^2}} \\ & \times & \sum_{\small \mbox{Levitating Paths}}    e^{- \frac i \hbar  z_{f}(\overrightarrow{\epsilon},t_N) (p-p_N)} \quad \,, \nonumber
\end{eqnarray}
The altitudes $z_{f}(\overrightarrow{\epsilon},t_N)$ are the end points of the trajectories
 followed by the atomic wave-packets of maximum upward momentum, i.e. which pick up exactly a quantum of upward momentum
 $\hbar k$ during each pair of copropagating light pulses. With a fine adjustment of the interpulse period $T$, these packets are maintained in suspension, theqrefore we refer to these atomic packets as the ``levitating packets'' and to their trajectories as the ``levitating paths''. We emphasize that the wave-function defined by Eq.\eqref{eq:sommeresonante1} is not normalized to unity, more precisely its square integral gives the levitating fraction $p_{ \mbox{lev}}(t_N)= \int {\rm d} p |\psi_{\mbox{lev}}(p,t_{N})|^2 $. At the time $t_N$, all the levitating wave-packets have acquired the same number of upward momentum quanta. However, these packets are located at different heights, depending on the times at which the successive momentum transfers happened. It is convenient to assign to each atomic wave-packet a $2N$-component vector $\overrightarrow{\epsilon}=(\epsilon_1,...,\epsilon_{2N})$ describing the history of these momentum transfers. We set $\epsilon_j=1(-1)$ if the wave-packet acquired a quantum of upward (downward) momentum at the $j_{\mbox{th}}$ pulse, and $\epsilon_j=0$ if the wave-packet did not go through a transition. The levitating atomic packets pick up exactly one quantum of upward momentum for each pair of copropagating light pulses, i.e. their vector $\overrightarrow{\epsilon}$ satisfies
 \begin{equation}
 \epsilon_{2n-1}+\epsilon_{2n}=1 \quad \mbox{for} \quad  n =1,...,N \,.
  \end{equation}
   We note $\cal{L}_{N}$ the set of such vectors. From the condition above, it is obvious that the vectors $\overrightarrow{\epsilon}$ associated with levitating
 end points satisfy $\forall j \in \{ 1, 2 N\} \: \epsilon_j \geq 0$.  The end point $z_{f}(\overrightarrow{\epsilon},t_N)$ can be expressed as a simple function of the vector $\overrightarrow{\epsilon}$:
\begin{eqnarray}
\label{eq:analytical zf}
& \: & z_{f}(\overrightarrow{\epsilon},t_N) =  z_0 - \frac 1 2 g t_N^2 \nonumber \\
& + &  \sum_{n=1}^{N} \left[ \frac {} {} \epsilon_{2n-1} v_r  \left( t_N- (n-1) T  \right)+\epsilon_{2n} v_r \left(t_N - n T \right)  \right] \nonumber
\end{eqnarray}
$v_r=\hbar k/m$ is the recoil velocity picked up by the wave-packet at each momentum transfer. For the levitating trajectories, one can simplify the expression above by using that $\overrightarrow{\epsilon} \in \cal{L}_N$, which gives
\begin{eqnarray}
\label{eq:analytical zf2}
 z_{f}(\overrightarrow{\epsilon},t_N)  = &  z_{0N} +  \sum_{n=1}^{N} \left( \epsilon_{2n-1}  - \epsilon_{2n} \right) v_r \frac {T} {2} \,.
\end{eqnarray}
with
\begin{equation}
z_{0N} =z_0  + \frac 1 2 N^2 (v_r-g T) T
\end{equation}

 In order to compute analytically the sum of Eq.\eqref{eq:sommeresonante1}, we express it as a function the vector $\overrightarrow{\epsilon}$. Using Eq.\eqref{eq:analytical zf2}, one obtains
  \begin{eqnarray}
\label{eq:analytical sum resonant2}
 \psi_{\mbox{lev}}(p,t_{N})  &  = &   \frac {e^{i \Phi_N}} {2^{N} \pi^{1/4} \sqrt {w_p} }  e^{-
	\frac {(p-p_N)^2} {w_p^2}} e^{- \frac {i} {\hbar} z_{0N}(p-p_N)}  \nonumber \\
& \times &      \sum_{\overrightarrow{\epsilon} \in \cal{L}_{N}}
e^{  \sum_{n=1}^{N} \left( \epsilon_{2n} - \epsilon_{2n-1}   \right)  \frac  {i v_r  T} {2 \hbar} (p-p_N) } \,.
\end{eqnarray}
The global phase $\phi_N$, depending on the phase convention chosen for the lasers, is not relevant to predict the fringe pattern of the interferometer. Thus we do not specify here its exact expression. We have used the property that any vector $\overrightarrow{\epsilon} \in \cal{L}_{N}$  satisfies $\sum_{j=1}^{2N} \epsilon_j=N$. The sum in the right hand side is analogous to an effective canonical partition function in which the configuration would be described by the vector $\overrightarrow{\epsilon}$. Since the coordinates satisfy $\epsilon_j \in \{0,1\}$, and since the pairs of values $(\epsilon_{2n_1},\epsilon_{2n_1+1})$ and $(\epsilon_{2n_2},\epsilon_{2n_2+1})$ are summed independently for $n_1 \neq n_2$, this sum factorizes.
The statistical analogue would be an assembly of double wells containing exactly one particle, the double wells being independent from one another. The two wells have here effective energies that are of opposite sign but equal in absolute value. The usual dimensionless quantity $\beta E$ is replaced by the complex number $\pm  v_r (p-p_N) (iT)/ \hbar$. Using the fact that each double well is independent, one can recast this effective partition function as
\begin{eqnarray}
\label{eq:analytical sum resonant2}
\psi_{\mbox{lev}}(p,t_{N})  & = &  \frac {1} { \pi^{1/4} \sqrt {w_p}} e^{i \Phi_N} e^{-
	\frac {(p-p_N)^2} {2 w_p^2}}    e^{- \frac {i} {\hbar} z_{0N}(p-p_N)}  \nonumber \\
& \times & \cos^N  \left( \frac  {v_r  T} {2 \hbar} (p-p_N)  \right) \,.
\end{eqnarray}
This expression shows that, as announced previously, the levitating wave-function experiences an exponential localization in momentum space as the number of pulses increases. This localization can be interpreted as a filtering in momentum operated by the multiple-wave interference. The broadening of the sample size along the vertical axis, which happens through the splitting of the levitating trajectories,
reflects the back-action of such momentum localization. The analogy between wave-function and partition function is reminiscent of the form of the propagator for atomic waves.

Using Eq.\eqref{eq:analytical sum resonant2}, one can evaluate the fraction of the initial cloud maintained in levitation after a $N$ pulse pairs.
\begin{equation}
p_{\mbox{lev}} \: = \:   F \left( \frac {k w_p} {2 m} T, 2N \right)
\end{equation}
with
\begin{equation}
F_1(\alpha,n)= \frac {1} { \sqrt{\pi} } \int_{-\infty}^{+\infty} {\rm d} u \:  e^{-
	u^2} \: \cos^{n}  \left(\alpha u  \right)
\end{equation}
Clearly $F(\alpha,n) \simeq 1$ only if $\alpha \ll 1$, which is the condition on the coherence length derived earlier.  This condition must be satisfied in order to obtain interferences between two atomic packets flying on different paths. However, we shall see that a certain sample fraction can still be suspended even if this condition is not met. Levitation happens then through constructive interferences at the intersection of the different paths, i.e. in the nodes of the suspended network.

 \section{Multiple-wave atomic levitation in the short pulse regime.}
 \label{sec:nonresonant short pulse regime}

In this Section, we compute the fringe pattern of the multiple-wave atom interferometer with a simplified model for the light pulses. Such model is valid when their duration $\tau$ is much shorter than the resonant period $T_0$, and when the pulses are nearly-resonant, i.e. detuned from a frequency $\delta \omega \ll 1 / \tau$, which is the regime considered for this setup. The light pulses, acting as temporal atomic beam splitters, can then be modeled through an instantaneous Rabi matrix with elements of equal norm. This model yields excellent agreement with the numerical simulations performed in the original proposal~\cite{Impens09b}. In the considered regime, the different levitating paths are equally populated, and the variation in the total levitating atomic population arises purely from interferences effects. In this sense, the short pulse limit is the paradigm of the levitation through multiple-wave atomic interference and corresponds to a perfect quantum trampoline. \\

\subsection{Clock fringes.}

\begin{figure}[htbp]
\begin{center}
\includegraphics[width=8.3cm]{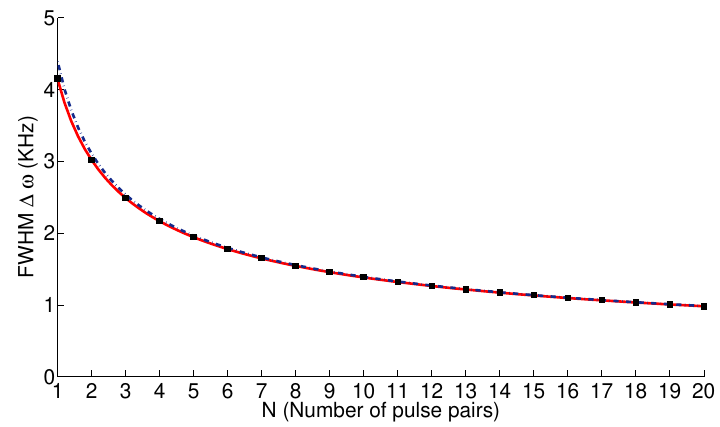}
\end{center}
\caption{Full width at half maximum (KHz) for the levitating population as a function of the number of pulse pairs.  Full red curve shows the bandwidth obtained with Eq.\eqref{eq:shortpulse clock bandwidth exact}. Blue dashed curve denotes the simplified expression of Eq.\eqref{eq:shortpulse clock bandwidth approximate}. The dots denote the bandwidth obtained from the numerical simulations of the levitating wave-packets used in Ref.\cite{Impens09b}.}
\label{fig:shortpulse bandwidth analytical simulation comparison}
\end{figure}

  Here we still assume that the interpulse-period $T$ is fixed at its resonant value $T:=T_0$, but the pulse frequencies $\omega_{j}$ are now shifted from their resonant values. As before, the pulse phases are synchronized in order to yield a relative phase $\phi_c=0$ for each pair [See Eq.\eqref{eq:laser pulse phases} of Sec.\ref{sec:twopulse}]. In order to understand how these frequency shifts affect the levitating wave-function, we consider again the scattering of a Gaussian lower-state wave-packet [See Eq.\eqref{eq:analytical packet resonant}], centered around the position $z_i$, on two upward-travelling light pulses interspaced by a duration $T_0$ and of respective frequencies $\omega_{1}=\omega_{1}^0+\delta \omega_{1}$ and $\omega_{2}=\omega_{2}^0+\delta \omega_{2}$. We know from Sec.\ref{sec:twopulse} that if the frequencies $\omega_{1,2}$ were resonant, the path $a$ and $b$ would yield an equal phase proportional to the invariant $I$ of Eq.\eqref{eq:twopulse expression invariant I}. Considering the expression for the laser phases in Eq.\eqref{eq:laser pulse phases} and the definition of the quantities $I_{a,b}$ in Eqs.(\ref{eq:twopulse Ia},\ref{eq:twopulse Ib}), one sees that the frequency shifts $\delta \omega_1$ and $\delta \omega_2$ induce a phase imbalance between the paths $a$ and $b$:
  \begin{eqnarray}
  \label{eq:clock analytical path imbalance}
 I_a(z_i,p_N,T_0)=I(k,p_N,T_0)- \hbar \delta \omega_{2}\frac {T_0} {2} \nonumber \\
 I_b(z_i,p_N,T_0)= I(k,p_N,T_0)+ \hbar \delta \omega_{1} \frac {T_0} {2} \,.
 \end{eqnarray}
Following the lines of Section~\ref{sec:twopulse}, the levitating wave-function  after the pair of upward-traveling pulses  is given by Eq.\eqref{eq:analytical resonant recursive twopacket} with the values above for $I_{a,b}$. Using the relations $z_{fa}=z_i - \hbar k T_0 /(2 m)$ and $z_{f b}=z_i + \hbar k T_0/(2 m)$, one obtains
\begin{eqnarray}
\label{eq:analytical nonresonant recursive twopacket up}
& \: & \psi_{\mbox{lev} \: b}(p,t_{N+1})  =  \frac {i e^{i \left( \Phi_N+I(k,p_N,T_0) / \hbar \right)}} {2^{N+1} \pi^{1/4} \sqrt {w_p}}
  \nonumber \\
 & \times & e^{- \frac {(p-p_{N+1})^2} {2 w_p^2}} e^{-i \frac {z_i} {\hbar} (p-p_N)} \\
& \times &  \sum_{\epsilon_1,\epsilon_2} e^{ i \left[  \epsilon_1 \left( \frac {} {} \delta \omega_{1}-  k(p-p_{N+1}) / m \right)- \epsilon_2 \left( \delta \omega_{2} -\frac {} {} k(p-p_{N+1}) / m  \right) \right]\frac { T_0} {2}} \nonumber
\end{eqnarray}
We sum only on the values $(\epsilon_1,\epsilon_2) \in \{ 0,1 \}^2$ such that $\epsilon_1+\epsilon_2=1$. One must also consider the scattering of a Gaussian upper-state wave-packet, centered initially around $z'_i$, on a pair of downward-traveling pulses separated by the duration $T_0$, and whose frequencies are noted $\omega_3$ and $\omega_4$. Since the gain of an upward momentum is then associated with a photon emission instead of a photon absorption, one simply needs to replace in the previous expression the frequencies $\omega_1$ and $\omega_2$ by the values $-\omega_3$ and $-\omega_4$ respectively. The result reads
\begin{eqnarray}
\label{eq:analytical nonresonant recursive twopacket down}
& \: & \psi_{\mbox{lev} \: a}(p,t_{N+2})  =  \frac {i e^{i \left( \Phi_{N+1}+I^{-}(k,p_{N+1},T_0) / \hbar \right)}} {2^{N+2} \pi^{1/4} \sqrt {w_p}}\nonumber \\
 & \times &  e^{- \frac {(p-p_{N+2})^2} {2 w_p^2} } e^{-i \frac {z'_i} {\hbar} (p-p_N)} \\
 & \times &  \sum_{\epsilon_3,\epsilon_4} e^{ i \left[  -\epsilon_3 \left( \frac {} {} \delta \omega_{3}+  k(p-p_{N+1}) / m \right)+ \epsilon_4 \left( \delta \omega_{4} +\frac {} {} k(p-p_{N+1}) / m  \right) \right]\frac { T_0} {2}} \nonumber
\end{eqnarray}
Similarly we sum on $(\epsilon_3,\epsilon_4) \in \{ 0,1 \}^2$ satisfying $\epsilon_3+\epsilon_4=1$. $I^{-}(k,p_{N+1},T_0)$ denotes the invariant associated with the sample propagation in a pair of resonant downward-travelling $\pi/2$-pulses. Its expression, not needed here, can be obtained by replacing the resonant frequencies $\omega_{1}^0$ and $\omega_2^0$ respectively with the values $-\omega_3^0=-\omega_{2}^0$ and $-\omega_4^0=-\omega_1^0$ in the expressions of $I_a$ and $I_b$ [Eqs.(\ref{eq:twopulse Ia},\ref{eq:twopulse Ib})]. Using the Eqs.(\ref{eq:analytical nonresonant recursive twopacket up},\ref{eq:analytical nonresonant recursive twopacket down}), one can show recursively that the wave-function after $N=2M$ pulse pairs is of the form
   \begin{eqnarray}
\label{eq:wavefunction N interferometers without gravity}
 \psi_{\mbox{lev} a}(p,t_{N})  & = & \frac {1} {2^N  \pi^{1/4} \sqrt {w_p}}  e^{i \Phi_N} e^{-(p-p_N)^2/ 2 w_p^2} \\
& \times & \sum_{\overrightarrow{\epsilon} \in \cal{L}_N}  C(\overrightarrow{\epsilon},\overrightarrow{\omega})   e^{- \frac i \hbar  z_{f}(\overrightarrow{\epsilon},t_N) (p-p_N)}  \nonumber
\end{eqnarray}
with
\begin{equation}
C(\overrightarrow{\epsilon},\overrightarrow{\omega})  =  \prod_{j=1}^{2 N}  e^{i (-1)^{\lfloor \frac {j-1} {2} \rfloor} (-1)^{j+1} \: \epsilon_j \delta \omega_j \frac {T_0} {2}} \nonumber
\end{equation}
 As in the previous Section, the factorization of this expression is easily obtained, either by using the former analogy of a two-particle partition function, either by using directly the recursive relations of Eqs.(\ref{eq:analytical nonresonant recursive twopacket up},\ref{eq:analytical nonresonant recursive twopacket down}):
   \begin{eqnarray}
\label{eq:wavefunction N interferometers without gravity}
&\: & \psi_{\mbox{lev}}(p,t_N)   =   \frac {e^{i \Phi_N}} {\pi^{1/4} \sqrt {w_p}} e^{-\frac {(p-p_N)^2} {2 w_p^2}}       \nonumber \\
& \times &  \prod_{n=0}^{M-1} \cos \left[ \left( \frac {\delta \omega_{4n+1}+\delta \omega_{4n+2}} {2}-  \frac {k(p-p_{N})} {m} \right) \frac {T_0} {2} \right]  \nonumber \\
    & \times & \cos \left[ \left( \frac {\delta \omega_{4n+3}+\delta \omega_{4n+4}} {2}+  \frac {k(p-p_{N})} {m} \right) \frac {T_0} {2} \right]
\end{eqnarray}
Again, we do not specify the global phase factor $\phi'_N$. We can now describe the clock fringes assuming that all pulses are equally detuned, i.e. $\forall \: i \: \delta \omega_i=\delta \omega$. The levitating population yields simply
\begin{equation}
p_{\mbox{lev}} \: = \:   F_{2} \left( \frac {k w_p} {2 m} T, N, \delta \omega \frac {T_0} {2} \right)
\end{equation}
with
\begin{equation}
F_2(\alpha,n, \phi)= \frac {1} { \sqrt{\pi} } \int_{-\infty}^{+\infty} {\rm d} u \:  e^{-
	u^2} \: \cos^{n}  \left(\alpha u  + \phi \right) \: \cos^{n}  \left(\alpha u  - \phi \right)
\end{equation}
We consider here the limit of long sample coherence length, which corresponds to $\alpha \ll 1$, and which is also the regime discussed in Ref.\cite{Impens09b}. In this limit, the levitating fraction becomes simply
\begin{equation}
\label{eq:shortpulse clock bandwidth exact}
p_{\mbox{lev}} \: = \:   \cos^{2 N}  \left( \delta \omega \frac {T_0} {2}\right)
\end{equation}
This gives readily the full width at half maximum (FWHM) $\Delta \omega$, yielding by definition $p_{\mbox{lev}}(\pm \frac 1 2 \Delta \omega )=1/2$, i.e. $
\Delta \omega = 4 \arccos \left( 1 / 2^{1/(2 N)}  \right) / T_0$. In the limit where $N \gg 1$, one easily sees that $\Delta \omega \sim T_0 / \sqrt{N}$. A Taylor expansion in the equation above yields $p_{\mbox{lev}}(\delta \omega) \simeq  e^{-  N \delta \omega^2 T_0^2 /4}$, hence
\begin{equation}
\label{eq:shortpulse clock bandwidth approximate}
\Delta \omega \simeq \frac {\sqrt{2} \log(2)} {T_0 \sqrt{N}} \,.
\end{equation}
One retrieves the usual improvement of the frequency resolution $\Delta \omega$ with the interrogation time $T_{\mbox{int}}=N T_0$  as $\Delta \omega \propto T_{\mbox{int}}^{-1/2}$, which is typical of atomic clocks. These expressions for the width of resonance are plotted in the Fig.\eqref{fig:shortpulse bandwidth analytical simulation comparison}. This figure shows that the width of resonance obtained in Eq.\eqref{eq:shortpulse clock bandwidth approximate} with a Taylor expansion is accurate even for a small number of pulses. Excellent agreement is obtained between the expression of Eq.\eqref{eq:shortpulse clock bandwidth exact} and the bandwidth computed from the simulations used in Ref.\cite{Impens09b}. As in Ref.\cite{Impens09b}, we considered the mass $m$ of $\:^{87} \mbox{Rb}$ atoms and a wave-vector $ k \simeq \: 10^7 m^{-1}$. 

\subsection{Gravimeter fringes.}

 When the setup operates as a gravimeter, one scans the interpulse period $T$ around its resonant value $T_0$. The frequencies are chosen as resonant for an average atomic momentum $p'_0$ at the beginning of the four-pulse sequence, i.e. $\omega_{4j+1}=\omega_{4j+4}=\omega_1^0$ and $\omega_{4j+2}=\omega_{4j+3}=\omega_2^0$, with $\omega_{1,2}^0$ satisfying Eq.\eqref{eq:twopulse resonance frequencies} with $p'_0$.  We note $p_j$ is the momentum of the levitating wave-packets at the the time $t_j=j T$. The mismatch in the resonance condition induces a drift in momentum, i.e. $p_j=p_0+ j  (\hbar k - m g T)$. One sees on the resonance condition~[Eq.\eqref{eq:twopulse resonance frequencies}] that the mismatch of the atomic momentum with respect to the resonant value $p'_0$ is responsible for the frequency detuning of the $j_{th}$ pulse
\begin{equation}
\delta \omega_{j}= (-1)^{\lfloor \frac {j+1} {2}  \rfloor} \frac {k} {m} \left[ p_0-p'_0+ \zeta(j)  (\hbar k - m g T) \frac {} {} \right] \,. \nonumber
\end{equation}
The first factor in the right hand side accounts for the reversal of the direction of the wave-vector $\mathbf{k}$ between each pair of pulses. We have introduced for convenience the auxiliary function $\zeta(j) T$ which gives the instant at which the $j_{\mbox{th}}$ pulse is performed in the considered sequence: $\zeta(4m+1)=2m$, $\zeta(4m+2)=2m+1$, $\zeta(4m+3)=2m+1$ and $\zeta(4m+4)=2m+2$. The previous discussion on the path imbalance induced by a detuning can be readily applied. One obtains a levitating wave-function given by Eq.\eqref{eq:wavefunction N interferometers without gravity} with the frequency detunings $\delta \omega_j$  replaced by the values of the equation above. In the limit of long condensate coherence length, the levitating fraction reads
\begin{equation}
\label{eq:wavefunction N interferometers without gravity}
p_{\mbox{lev}} =    \prod_{j=1}^{2 M}  \cos^2 \left( \frac {(p_0-p_0')} {2} +\frac {j} {2} (k g \delta T)  T_0 \right) \,. \nonumber
\end{equation}
Losses in the levitating cloud arise from the initial momentum shift and from the mismatch in the resonant period. The gravitational acceleration $g$ is determined from Eq.\eqref{eq:levitation resonance T}. One assumes that the recoil velocity is known with an excellent precision, so that the precision of the measurement of the acceleration $g$ is simply limited by the determination of $T_0$, i.e. $\Delta g/g=\Delta T_0 /T_0$. Considering that a fractional loss $\epsilon$ can be detected in the levitating atomic population, one can determine the resonant period $T_0$ up to an uncertainty $\Delta T_0 = T_2-T_1$, with $T_1$ and $T_2$ such that $p_a(T_1)=p_a(T_2)=1-\epsilon$. Taking the logarithm of Eq.\eqref{eq:wavefunction N interferometers without gravity}, using a Taylor expansion and replacing the sum by an integral, one finds
\begin{equation}
\label{eq:sensibilite gravimetre}
\frac {\Delta g} {g} \simeq \frac {\sqrt{- 6 \log(1-\epsilon})} {k \: g \: T_0^2} \frac {1} {N^{3/2}} \,.
\end{equation}
As in Ref.\cite{Impens06,Hughes09}, the atomic velocity drift over the successive cycles yields a measurement accuracy scaling with the number $N$ of cycles as $\Delta g / g \propto 1/N^{3/2}$. The accuracy given above agrees very well with the accuracy predicted by numerical simulations of the sample. Both are plotted on the following figure.
\begin{figure}[htbp]
\begin{center}
\includegraphics[width=7cm]{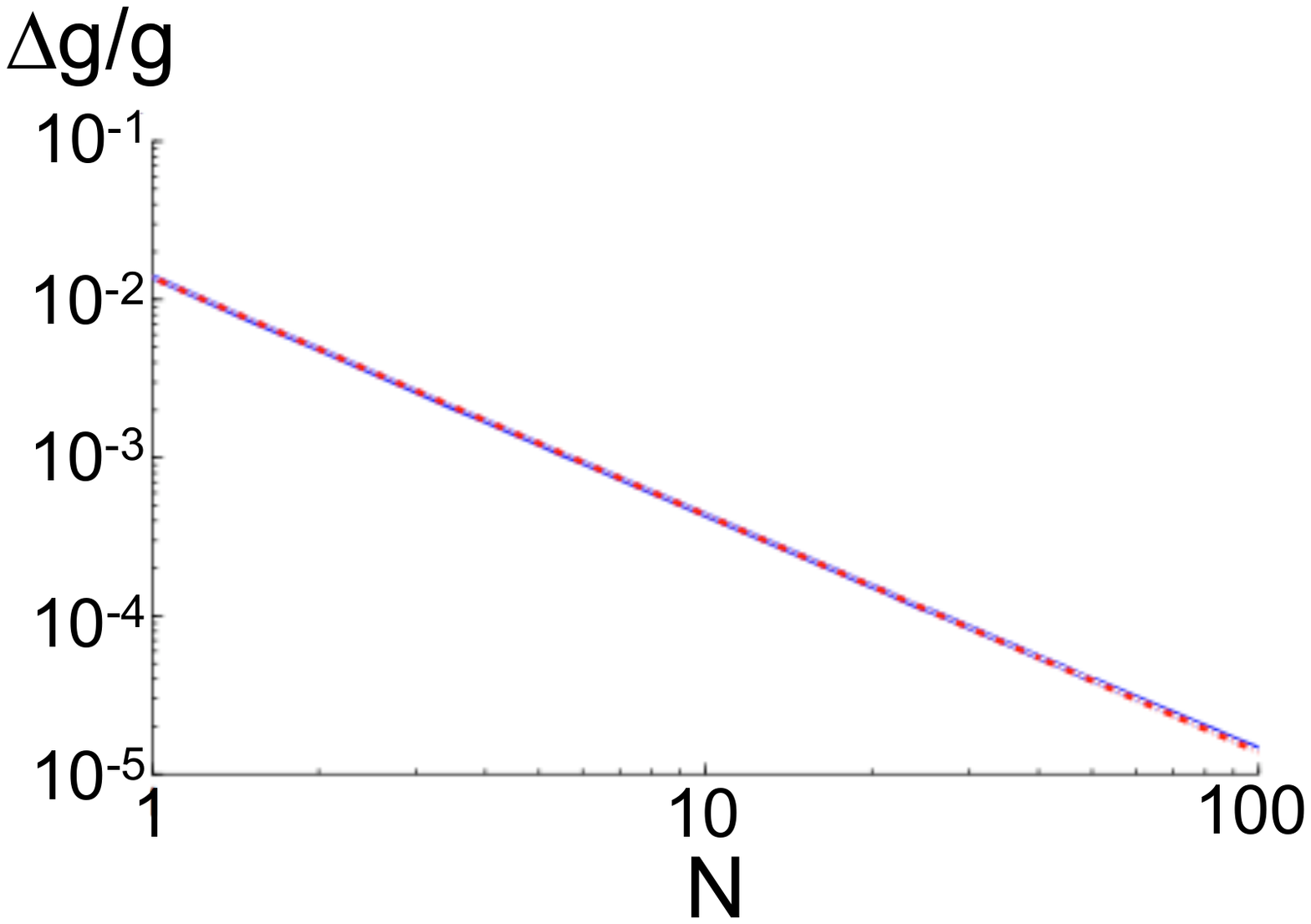}
\end{center}
\caption{Relative uncertainty on the gravitational acceleration predicted by Eq.\eqref{eq:sensibilite gravimetre} (red dashed curve) and  that obtained with numerical simulations (blue curve) as a function of the number of cycles $N$ (corresponding to 2N pulse pairs).}
\label{fig:shortpulse gravwidth}
\end{figure}

\section{Experimental realization of the setup.}
\label{sec:clock systems}

We investigate here suitable atomic species for the considered scheme. The considered two-level atom should have a clock transition whose life-time should fulfill two seemingly contradictory requirements. In order to interrogate the atoms over several bounces, this life-time should be much longer than the interpulse period $T_0$. It should also be short enough as to enable the realization of a $\pi/2$-light pulse in a duration less than $T_0/2$ with a realistic laser intensity. These two criteria imply a non-trivial trade-off between sharpness linewidth and coupling on the transition.

Besides, the atomic sample should be cold enough so that the short $\pi/2$ light pulses address its entire velocity width. In order to preserve the full atomic population in levitation, one should use an atomic sample with a coherence length much longer than the separation between two paths. As discussed here-after, this requirement is not met by fermionic samples: a Bose-Einstein condensate is required to obtain such global interferences. Nevertheless, partial levitation can still be achieved with fermions for a long duration.

\subsection{Local interferences with Ultra-cold Fermions.}

The compromise on the clock transition is ideally achieved by the $\:^{1} \mbox{S}_0- \:^{3} \mbox{P}_0$ transition of the fermionic isotope $\:^{171}\mbox{Yb}$~\cite{Ybtransitionpaper}, which has a wave-length $\lambda=578.4 \:  \mbox{nm}$ and a frequency width of $10 \: \mbox{mHz}$. For this transition, a $\pi/2$ pulse of $4 \: \mu \mbox{s}$ (on the order of $T_0/100$) can be realized with a laser power of $0.3 \: \mbox{W}$ focalized on a spot of radius $1 \: \mbox{mm}$ ($1/e^2$ beam radius)~\cite{Ybtransitionpaper}.

\begin{figure}[htbp]
\begin{center}
\includegraphics[width=8.3cm]{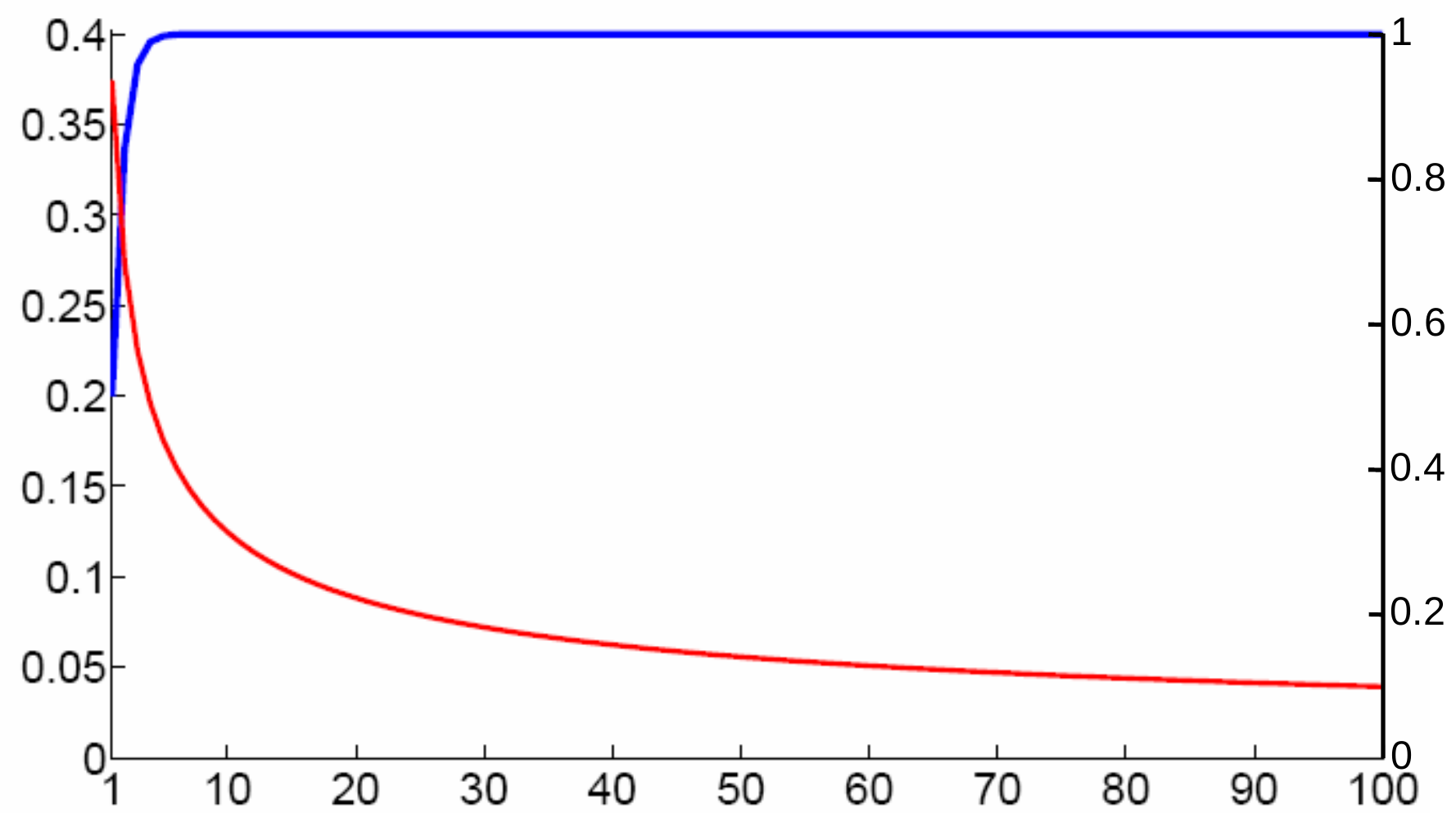}
\end{center}
\caption{Levitating fraction (red curve, left scale) and contrast (blue curve, right scale) as a function of the number of pulse pairs for a sample of $Yb$ atoms at a temperature $\Theta=130 \: \mbox {nK}$.}
\label{fig:shortpulse levitation and contrast}
\end{figure}

 Even if temperatures as low as $\Theta=130 \: \mbox {nK}$ have been achieved experimentally by sympathetic cooling with a cloud of $10^4$ atoms of this isotope~\cite{YbFroid}, well within the regime of quantum degeneracy, the associated dispersion in momentum is unfortunately too wide to obtain interferences between atomic packets propagating on different paths. Indeed, the coherence length $w$  is still much smaller than the distance $L$ between two adjacent paths, i.e. $L/w=\hbar k^2 \sqrt{k_B \Theta}/ m^{3/2} \sim 10$ with $k_B$ the Boltzmann constant. Nevertheless, the interferences of the wave-packets at the nodes of the network still enable to levitate a significant fraction of the atomic population and yield a sufficient contrast. Indeed, one can tailor either constructive, either destructive interferences at these nodes. We note that the interferences at each node correspond to those of an atomic Bord\'e-Ramsey interferometer issued from a node located at the same height and occurring at a time $2T$ earlier. When $(I_b-I_a) / \hbar \equiv 0 \: (\pi)$, the arms of the Bord\'e-Ramsey interferometer yield the same phase and thus constructive interferences at each node. In contrast, when $(I_b-I_a)/\hbar \equiv \pi/2 \: (\pi)$ , one obtains a phase detuning of $\pi$ between the arms of the Ramsey-Bord\'e interferometers yielding destructive interferences at each node where several trajectories intersect. These interferences quickly shut off the levitation as the number of pulse pairs increases: beyond ten pairs of pulse, the levitating fraction is found below $10^{-6}$, so that no atom in the sample is expected to levitate. The contrast is defined as $C=(p_{\mbox{lev}}(0)-p_{\mbox{lev}}(\pi/2))/(p_{\mbox{lev}}(0)+p_{\mbox{lev}}(\pi/2))$, where $p_{\mbox{lev}}(0)$ and $p_{\mbox{lev}}(\pi/2)$ are the levitating fraction for constructive and destructive interferences respectively. Fig.\ref{fig:shortpulse levitation and contrast} gives the levitating fraction at resonance as well as the contrast of the interference pattern for a sample of ultra-cold $\:^{171}\mbox{Yb}$ atoms. The resonant period is
 $T_{0 \: \:^{171}\mbox{Yb}} \simeq 0.4 \: \mbox{ms}$ for this isotope. Starting with a sample of $10^4$ atoms, one expects to maintain roughly $400$ atoms in suspension after $100$ pairs of pulses which is still enough to be detected with state-of-the-art techniques.

\subsection{Global interferences with ultra-cold Bosons.}

In order to obtain interferences between atomic packets flying on different paths, the initial sample must have a sufficient coherence length. Such coherence is only achieved by BECs, so that this time bosonic species need to be considered. In order for the setup to operate as a clock, the considered atom should own a narrow transition in its internal structure. For this purpose, one can consider using the forbidden optical transition $\:^{1} \mbox{S}_0- \:^{3} \mbox{P}_0$, enhanced by a dc magnetic field, of neutral even isotopes of Yb, Sr, Mg, or Ca~\cite{Hollberg06}. In the following discussion, we use the atomic transition parameters presented in this reference.

We focus on the two species $\:^{174} \mbox{Yb}$ and $\:^{40} \mbox{Ca}$, which have been Bose-condensed recently~\cite{YbCaBEC}, and which have different advantages. Since both the recoil velocity and the resonant period $T_0$ decrease with the atomic mass, the requirement on the coherence length is less stringent for heavy atoms. This point advocates for the use of heavy ultra-cold samples such as the $\mbox{Yb}$ isotopes, for which the coherence length required to obtain inter-path interferences is roughly $w_{\mbox{Yb}} \simeq 1 \: \mu \mbox{m}$, to be compared with $w_{\mbox{Ca}} \simeq 10 \: \mu \mbox{m}$ for Ca atoms. On the other hand, since $T_{0 \: \mbox{Ca}} \gg T_{0 \: \mbox{Yb}}$, a setup using $\:^{40} \mbox{Ca}$ atoms involves a smaller Rabi pulsation on the clock transition. This property benefits to Ca atoms, since a smaller Rabi pulsation requires less intense magnetic and light field to perform the transition, and thus smaller spurious frequency shifts associated with the second-order Zeeman effect and the Stark shift respectively. We impose a pulse duration $\tau= \: 0.3 T_0$ for the $\pi/2$-light pulses, which corresponds to an operation of the multiple-wave interferometer in the long-pulse regime, more precisely with $60 \%$ of illumination time. We assume that the dc magnetic field $B$ can be controlled up to a relative accuracy of $10^{-4}$, while the light intensity $I$ is monitored with a relative accuracy of $10^{-4}$. Using $\:^{174} \mbox{Yb}$ atoms, one performs the desired light pulses by applying a magnetic field $B = 200 \: \mbox{G}$ and a laser power $P \simeq 50 \: \mbox{mW}$ focused as above on a spot of radius $0.1  \: \mbox{mm}$, yielding a magnetic frequency uncertainty $\Delta_{B \: \mbox{Yb}}=0.50 \: \mbox{Hz}$ and a light shift uncertainty $\Delta_{L \: \mbox{Yb}}=0.46 \: \mbox{Hz}$. With $\:^{40} \mbox{Ca}$ atoms, one realizes instead the pulse with a smaller magnetic field $B = 40 \: \mbox{G}$ and a laser power $P \simeq 120 \: \mbox{mW}$ focused on a similar spot, yielding the respective magnetic and light shift uncertainties $\Delta_{B \: \mbox{Ca}} \simeq \Delta_{L \: \mbox{Ca}} \simeq 0.27 \: \mbox{Hz}$. 
These figures can seem high in comparison with other schemes performing an interrogation with a Rabi pulsation in the $\mbox{Hz}$ range such as optical network setups. These significant shifts and uncertainties are caused by the short pulse constraint imposed by the value of the resonant period $T_0$. This constraint would be largely relaxed when operating this architecture with an atomic transition in the UV range~\cite{UVclock}: the period $T_0= (\hbar/ m c) \times (\omega / g)$ would become much longer, thereby enabling atomic levitation with weak laser fields inducing smaller frequency shifts.

\section{Conclusion.}

It is appropriate to compare briefly the theoretical method used here with the treatment of other recent atomic levitation experiments~\cite{Hughes09,SaintVincent10}. These experiments rely on the illumination with standing light waves instead of the traveling waves considered here.

In all the atomic levitation experiments~\cite{Impens06,Impens09b,Hughes09,SaintVincent10}, the analysis of the sample sustentation relies on the articulation between two distinct regimes of propagation, namely the atomic propagation in the dark and the atomic propagation in the laser field. The propagation in the dark has been solved exactly with the atom optical $ABCD$ formalism~\cite{BordeABCD} under the following assumptions: the atoms evolve in an external potential which is quadratic, and atomic interactions are sufficiently weak to be neglected. Such assumptions are legitimate for typical clock experiments. In the absence of illumination, atoms are subject only to
the local gravitational potential which is very well-approximated by a quadratic potential. Mean-field effects can be included in the $ABCD$ method for moderate interactions when a single atomic cloud is involved~\cite{Impens09ABCD}, or if the system is loaded with an atom laser~\cite{AtomLaserABCD}. So far our treatment is equivalent to that used in the Refs.\cite{Hughes09,SaintVincent10}.

 However, a clear distinction needs to be made between our proposal and the experiments~\cite{Hughes09,SaintVincent10} regarding the propagation in the laser field. During the sample illumination, the light field couples plane atomic waves of different momenta. In our proposals involving traveling waves, the light field couples each atomic momentum eigenstate $|\mathbf{p} \rangle$ to a single momentum eigenstate $| \mathbf{p} + \hbar \mathbf{k} \rangle$ as expected from the atomic recoil. Such scattering, involving a single quantum channel, can be effectively modeled by means of a $2 \times 2$ Rabi matrix. In contrast, the scattering resulting from the illumination with a standing wave involve multiple quantum channels and couple instead each atomic momentum eigenstate $| \mathbf{p} \rangle$ to a multiplicity of states $\{ |\mathbf{p}+ m \hbar \mathbf{k} \rangle, m \in \mathbb{Z} \}$. The description of the scattering is thus much more complex when standing light waves are involved instead of traveling waves. It is not clear whether a similar summation with an effective partition function can be obtained when standing waves are used. One should note, however, that in practice for the experiments reported in the Refs.\cite{Hughes09,SaintVincent10}, the pulses are tailored so that the standing wave scattering involve predominantly the first few orders - in Ref.\cite{SaintVincent10}, scattering beyond the second order can be safely neglected -. A finite-dimensional Rabi matrix is then also suitable do describe the atomic scattering by standing light waves.

To summarize, we have developed a theory of atom interferometers using multiple-wave levitation. We have used a propagation invariant to describe the atomic scattering by a periodic series of short light pulses. The sensitivity of the sensor as an atomic clock or as a gravimeter has been computed, and yields excellent agreement with the numerical simulations presented in Ref.\cite{Impens09b}. We have proposed to use either fermionic or bosonic species to explore localized or global interferences in the levitating atomic sample. Using recent advances in the cooling and spectroscopy of Yb and Ca isotopes, we have shown that such clock system can be implemented with state-of-the art experimental techniques. Further developments will include a more sophisticated treatment of atomic beam-splitters in order to explore the system behavior in the long pulse regime. Last, we mention that our scheme would be particularily well-suited for clocks running with high-frequency transitions in the $UV$ range~\cite{UVclock}.

\section*{Acknowledgements.}

The authors would like to thank Luiz Davidovich, Arnaud Landragin, Yann Le~Coq, Paulo Americo Maia-Neto, Tanja Mehlsta\"{u}bler, Peter Wolf and Nicim Zagury for interesting discussions. FI acknowledges partial support from Ecole Polytechnique (Postdoctoral Fellowship) and from Minist\`ere des Affaires Etrang\`eres (Lavoisier-Br\'esil Fellowship).\\

\section*{Appendix: Expression of the levitating wave-packet.}

We establish by recursion the expression of Eq.\eqref{eq:sommeresonante1}:
\begin{eqnarray}
\psi_{\mbox{lev}}(p,t_{N}) & = & \frac {1} {2^{N} \pi^{1/4} \sqrt {w_p}} e^{i \Phi_N} e^{-
\frac {(p-p_N)^2} {2 w_p^2}} \nonumber \\ & \times & \sum_{\small \mbox{Levitating Paths}}    e^{- \frac i \hbar  z_{f}(\overrightarrow{\epsilon},t_N) (p-p_N)} \quad \,. \nonumber
\end{eqnarray}
 For $N=0$, the sum in this equation contains a single term which is the initial lower-state Gaussian wave-packet expressed in the momentum picture. This proves the desired property at the rank $N=0$. Let ut assume that this property is satisfied at the rank $N$. To establish it at the rank $N+1$, we consider the propagation of the wave-function of Eq.\eqref{eq:sommeresonante1} in a sequence of two light pulses of short duration, of respective initial instants $t_N=N T$ and $t_{N+1}=(N+1)T$. We assume that the levitating wave-packet is in the lower state at the instant $t=t_N$, and that the light pulses are upward travelling.  Since the propagation is linear, one can consider independently each Gaussian wave-packet present in the sum of Eq.\eqref{eq:sommeresonante1}. It is thus sufficient to prove the two following facts. Any Gaussian packet in the sum of  Eq.\eqref{eq:sommeresonante1}, associated with the altitude $z_i$, is split into two levitating wave-packets, whose central altitudes $z_{fa}$ and $z_{fb}$ are obtained by following either the lower or the upper arche issued from the altitude $z_i$ [See Fig.\ref{fig:sequence deux pisurdeux}]. Besides, these two wave-packets, written in the form of Eq.\eqref{eq:sommeresonante1}, bear a phase $\phi_{N+1}$ which is independent of the altitude $z_i$ of the initial wave-packet.

Following these lines, we consider the following wave-function at the instant $t_N$
\begin{equation}
\label{eq:analytical packet resonant}
\psi_{\mbox{lev}}(p,t_N)= \frac {e^{i \Phi_N}} {2^N \pi^{1/4} \sqrt {w_p}} \left( \begin{array} {c} 0 \\ e^{-
 \frac {(p-p_N)^2 } {2 w_p^2} } e^{-  \frac {i} {\hbar} z_i (p-p_N) } \end{array}  \right) \,,
\end{equation}
The lower-state wave-function is a Gaussian wave-packet in the momentum picture, which would be noted $wp(p,z_i,p_N,w,t_N)$  with the conventions of Eq.\eqref{eq:twopulse initial wave packet 1D Gaussian position}. To compute its propagation between the instants $t_N$ and $t_{N+1}$, one can apply the discussion of Section~\ref{sec:twopulse}, which gives readily the levitating wave-packet at time $t_{N+1}$. We discard here the lower-state component of Eq.\eqref{eq:twopulse final momentum wavepacket}, since the corresponding wave-packets have missed one quantum of upward momentum and thus do not contribute to the levitating wave-function. Using Eqs.(\ref{eq:twopulse phases},\ref{eq:twopulse invariant I}), this wave-function can be expressed with the path invariants noted $I_a=I_a(z_i,p_N,T)$ and $I_b=I_b(z_i,p_N,T)$
\begin{eqnarray}
\label{eq:analytical resonant recursive twopacket}
& \: & \psi_{\mbox{lev} \: b}(p,t_{N+1})  =  \frac {i e^{i \Phi_N}} {2^{N+1} \pi^{1/4} \sqrt {w_p}}
 e^{- \frac {(p-p_{N+1})^2} {2 w_p^2} } \nonumber \\
 & \times &  \left( e^{- \frac i \hbar z_{f a} (p-p_{N+1})+i \frac {I_a} {\hbar}}+ e^{- \frac i \hbar z_{f b} (p-p_{N+1})+i \frac {I_b} {\hbar}} \right) \nonumber \\
 \label{eq:analytical resonant philev}
\end{eqnarray}
 $p_{N+1}=p_N+\hbar k -m g T$ is the momentum of the levitating packets at time $t=t_{N+1}$. As outlined in Section \ref{sec:twopulse}, when the pulse frequencies satisfy the resonance conditions, the quantities $I_a$ and $I_b$ are equal and given by Eq.\eqref{eq:twopulse expression invariant I}, i.e. $I_a=I_b=I(k,p_N,T)$. One then obtains a phase factor $\phi_{N+1}=\phi_N+I(k,p_N,T)+\pi/2$ which is independent of the altitude $z_i$, and consequently equal for all the wave-packets present in the sum of Eq.\eqref{eq:sommeresonante1}. Besides, when the altitude $z_i$ takes all possible values of the end points $z_f$ of the levitating trajectories at time $t=t_N$, one obtains with $z_{fa}$ and $z_{fb}$ all the possible end points at the time $t=t_{N+1}$. The desired property is thus proven at the rank $N+1$. Should we have considered instead the transfer of atomic wave-packets from the upper to the lower state with downward-travelling pulses, the argument would have been naturally identical.\\

\end{document}